%Paper: hep-ph/9310323
%From: ARTRU@frcpn11.in2p3.fr
%Date: Wed, 20 Oct 93 18:37:23 MET

% INSTRUCTION FOR PRINTING THE FIGURES :
% Figures 1,2,3 are in the postscript file appended after the \end
% statement. This file is named "figprot.ps" in the TeX file.

\magnification\magstep 1
\font\xsl=cmbxsl10
%\nopagenumbers
\noindent
{\bf I.P.N.-Lyon \hfill LYCEN/9353\break
\null\hfill October 1993}

\vskip 3 true cm

\noindent
\xsl Proposals for measuring transversity distributions in deep inelastic
electron scattering and a model for E-704 asymmetries
\rm\footnote {$^1$} {\it Contribution to the V$^{th}$ International Workshop
on High Energy Spin Physics, \hfill\break
{}~~SPIN-93, September 20-24, 1993, Protvino, Russia}

\vskip 1.5 true cm
\centerline{Xavier ARTRU}
\medskip
{\it Institut de Physique Nucl\'eaire de Lyon, IN2P3-CNRS et
Universit\'e Claude Bernard,
\item{}43 Bd du 11 Novembre 1918, F-69622 Villeurbanne Cedex, France} \hfill

\vskip 3 true cm
\noindent
{\bf Abstract}
\vskip 0.5 true cm

We present the principles of the measurement of the quark
transversity distributions in semi-inclusive deep inelastic electron
scattering, which form the basis of HELP and one of the ELFE proposals.
A string model for Collins-type asymmetry in polarized quark fragmentation
function is proposed. A possible role of the Collins effect in the single
spin asymmetries observed by experiment E 704 at Fermilab is suggested.

\vfill\eject

\bigskip
\noindent
{\bf 1. Introduction.} At leading twist, the most complete information about
the short distance quark structure of the nucleon involves three distribution
functions : \par
\smallskip
\noindent
$\Delta q(x)= q_+(x) - q_-(x)$ in a nucleon of helicity $+{1\over2}$,
\hfill (1)\par
\smallskip
\noindent
$\Delta_T q(x)= q_{\bf\hat n}(x) - q_{-{\bf\hat n}}(x)$ in a nucleon of
transversity ${+{\bf\hat n}}$, \hfill (2)\par
\smallskip
\noindent
$q(x)=q_+(x) + q_-(x)=q_{\bf\hat n}(x) + q_{-{\bf\hat n}}(x)$. \hfill (3)

\smallskip
Recent EMC and SLAC experiments have measured
$g_1(x)={1\over 2} \sum_f e_f^2 \Delta q_f(x)$ but so far no experiment has
measured $\Delta_T q(x)$ (also called$^{1-3}$
$h_T(x)$, $\Delta_1 q(x)$ or $h_1(x)$).

A transversity state is obtained by superposition of two helicity states :
$$\vert{\bf\hat n}>=2^{-1/2}\,(\,\vert +> + \exp(i\phi)\, \vert ->\,)
\,,\eqno(4)$$
where $\phi$ is the azimuth of ${\bf\hat n}$
(this definition applies to massless particles as well).
As a general rule, transversity distributions or transverse spin asymmetries
can be considered as interferences between helicity amplitudes differing by
one unit of helicity. In particular, $\Delta_T q(x)$ is proportional to the
discontinuity of the helicity-flip, forward quark-nucleon scattering amplitude
(lower blob of Fig. 1a).

{\it A priori}, $\Delta_T q(x)$ differs from $\Delta q(x)$ (except in the
nonrelativistic quark model). They do not have the same evolution with $Q^2$.
Gluon distributions do not mix with $\Delta_T q(x)$.
In a simple covariant (quark + scalar diquark) model of the nucleon,
$\Delta_T q(x)$ is always positive, whereas $\Delta_T q(x)$ is negative for
large enough average intrinsic transverse momentum$^2$.
Thus $\Delta_T q(x)$ and $\Delta q(x)$ contain nonredundant informations about
nucleon structure.

\medskip
\noindent
{\bf 2. Helicity selection rules for transversity measurement.}
If one wants to measure $\Delta_T q(x)$ in Deep Inelastic
Lepton Scattering (DILS) in a fully inclusive way (unitarity diagram of
Fig. 1a), one is impeded by the approximate
conservation of quark helicity at the electromagnetic vertices.
This is not the case in doubly polarized Drell-Yan reaction$^{1,2}$ (Fig. 1b),
which measures
$\Delta_T q(x)$ in one proton $\times\, \Delta_T  {\bar q}(x')$ in the other
proton. We can generalize these results in a "selection rule"$^2$ for reactions
probing transverse spin at leading twist :
{\it in the unitarity diagram, the line of a transversely polarized quark
must connect two different hadrons}. A list of such reactions
is given in Ref$^{\,4}$. \par
A doubly polarized Drell-Yan experiment has been proposed for RICH.

\medskip
\noindent
{\bf 3. Semi-inclusive hyperon production in DILS.}
Another probe is obtained from the Drell-Yan one by {\it crossing}. The
prototype of it is$^5$
$$e + \uparrow N \to e' + \uparrow \Lambda + X \,, \eqno(5)$$
where the $\Lambda$ is emitted in the current fragmentation region.
A typical exclusive diagram is shown in Fig. 2
(we choose the $\Lambda$ because it is the simplest self-analysing hadron
but any hyperon could do the job). $\Delta_T \bar q(x')$ is crossed into
the {\it transversely polarized fragmentation function}
$\Delta_T\,f_{q\to\Lambda}(z)$.
The polarization of the $\Lambda$ is given by
$$\vec{\cal P}_{\perp\Lambda}={\cal R}_{N\Lambda}\  \vec{\cal P}_{\perp N}
\,\times\hat D_{\rm NN}(\hat\theta)\times
{\sum e_f^2\, \Delta_T\, q_f(x)\, \Delta_T\,f_{q_f\to\Lambda}(z)
\over \sum e_f^2 \,q_f(x)\, f_{q_f\to\Lambda}(z)}
\qquad (f=u,d,s,\bar u, \bar d, \bar s) \,,\eqno(6)$$
where ${\cal R}_{N\Lambda}$ is the rotation
about the normal to the scattering plane which
brings the initial nucleon momentum along the $\Lambda$ one, in the c.m. frame.

$$\hat D_{\rm
NN}(\hat\theta)={4\,(1+\cos\hat\theta)\over4+(1+\cos\hat\theta)^2}
= -2\,\hat s \hat u / (\hat s^2 + \hat u^2)
= 2\, E_e\, E'_e / (E_e^2 + {E'}_e^2) \eqno(7)$$
is the depolarization parameter of electron-quark scattering.

\medskip
\noindent
{\bf 4. Mesonic polarimeter.}
In the preceding reaction, the $\uparrow$ hyperon can be considered as as
{\it polarimeter} for the scattered quark. But quark fragmentation into hyperon
is rather rare, therefore a high number of DILS events is required.
Other types of quark polarimeter have been proposed by Nachtmann, Efremov,
Collins, which use fragmentation into meson. The Collins one$^6$ seems the most
appropriate to quark transversity. The fragmentation of a quark of
polarization $\vec{\cal P}$ should have an azimuthal dependance given by
$$f(z,\vec p_T,\vec{\cal P})= f(z,p^2_T) + \Delta_T f(z,p^2_T)
\times{\cal P}_{\perp} \times
\sin\left(\,{\rm azimuth}(\vec{\cal P}) - {\rm azimuth}(\vec p)\,\right)
\,.\eqno(8)$$
The tensor-polarized fragmentation function $\hat h_{\bar 1}(z)$ into $\rho$
meson, introduced by Xiangdong Ji$^7$, gives rise to such asymmetry.

\medskip
\noindent
{\bf 5. The ELFE$^8$ and HELP$^9$ proposals.}
Two experiments measuring $\Delta_T\, q(x)$ in semi-inclusive DILS have been
proposed :\par
\noindent
1) $e + \uparrow N \to e' + \uparrow \Lambda + X $,
at the European Laboratory For Electron (ELFE) which is in project.
Beam energy : 15 GeV ; target : polarized $^6$Li~d or $^6$Li~H.
A $4\pi$ detector is envisaged. To fix the ideas,
for an integrated luminosity of $2.5 \times 10^{40}$ cm$^{-2}$, one expects
about $10^5$ lambdas with $x_F>0$, $z>0.3$ and $<Q^2>=2.0\,{\rm GeV}^2$.
The $\Lambda$ polarization could reach 0.06 at $x_{Bj}\simeq 0.2$ and be
measured with a precision of 0.01.

\noindent
2) The HELP collaboration proposes to install a hydrogen jet target,
90 \% polarized, in the LEP tunnel ($E_e = 90$ GeV) in a "parasitic" mode.
Target thickness : $10^{13}$ atoms/cm$^2$.
In 6 months, about $10^4$ lambda should be collected ($2.2\,10^3$) for
$x_{Bj}>0.15$). In addition, the azimuthal asymmetry
in $e \,+ \uparrow N \to e' + {\rm meson} + X $ (Collins effect)
will be investigated, which increases the number of interesting events by more
than $10^2$.

Such experiments can measure $\Delta_T q(x)$ and $\Delta_T f(z)$
(or $\Delta_T f(z,k^2_T)$ of Eq.(8)) up to normalization factors, since only
the product of the two functions appear. Other probes will be necessary to
"calibrate" these functions. However, a nonzero signal would be by itself
of great significance.

\medskip
\noindent
{\bf 6. A string model for the Collins effect.}
Fig. 4 represent the string pulled in the $\hat z$ direction
by a fast quark $q_0$,
which fragments into a leading pion plus a string remnant. This quark
is supposed to have transverse polarization directed toward us
($s_y(q_0) = + {1\over 2}$).
The $q_1\bar q_1$ pair created during string breaking is assumed to be in
$S=1,\ J^P=0^+$ state (the $^3P_0$ model$^{10}$).  $q_0$ and $\bar q_1$
must have antiparallel spin in order to form a pion, therefore
$s_y(q_1)= s_y(\bar q_1)= - {1\over 2}$ and $L_y(q_1\bar q_1) = +1$.
Adopting an idea of the Lund group$^{11}$, this orbital momentum
is generated by the tunneling mechanism of quark pair creation :
$$L_y = +\, \kappa^{-1}\,[m_q^2 + p_T^2(\bar q_1)]^{1/2} \times p_x(\bar q_1)
\,,\eqno(9)$$
where $\kappa$ is the string tension. Furthermore,
$p_T({\rm pion}) = p_T(\bar q_1)$ ($= - p_T(q_1)$). Thus $p_x$ of the pion is
positive ($\pm$ some quantum uncertainty).

\medskip
\noindent
{\bf 7. Is Collins effect responsible for single spin asymmetry in E 704 ?}\par
\noindent
In $\uparrow p + N \to \pi + X$ at high $x_F$, the pion comes more likely
from the fragmentation of a valence quark of the proton. If we assume that this
quark is polarized like in the $SU(6)$ model and that a string-type
Collins effect takes place,
we recover$^{12}$ the experimental$^{13}$ signs of the single spin asymmetries
:
for upward proton polarization, $\pi^+$ and $\pi^0$ are preferentially
emitted to the left and $\pi^-$ to the right.
[ For $p_T({\rm pion}) \ge$ 2 GeV/c, the valence quark most probably
undergoes hard scattering. This results in a reduction of
the quark polarization by a factor $\hat D_{\rm NN}(\hat\theta)$ given by (6)
and a smearing of the Collins effect by integration over
the hard scattering angle. However,
$\hat\theta$ is not very large so that $\hat D_{\rm NN}(\hat\theta)$ is close
to unity. As regards smearing, it is not too severe,
due to the "trigger bias" effect. ]

\medskip
\noindent
{\bf 8. Conclusion.}
Quark transversity is a novel an promising observable for the understanding of
hadronic wave function in terms of bare quarks. It is a challenge to make its
first measurement. Among various probes, semi-inclusive Deep Inelastic Lepton
Scattering is one of the best candidate. The polarization of the scattered
quark
has to be detected either in the fragmentation into a hyperon or by azimuthal
asymmetry of a leading pion (Collins effect). An example of Collins effect is
provided by the string model. The later effect may also explain the single spin
asymmetries observed in $\uparrow p + N \to \pi + X$ at high $x_F$. If it is
the case, the quark transversity distribution in the proton should be large.

\medskip
\noindent
{\bf References}

\noindent
1) J.P. Ralston, D.E. Soper, Nucl. Phys. B 152, 109 (1979).

\noindent
2) X. Artru, M. Mekhfi: Z. Phys. C 45, 669 (1990).

\noindent
3) R.L. Jaffe, Xiangdong Ji, Nucl. Phys. B 375, 527 (1992) ;

\noindent
4) X. Artru, 10$^{th}$ Int. Symp. on High
Energy Spin Physics, Nagoya, Nov. 9-14, 1992.

\noindent
5) X. Artru, M. Mekhfi, Nucl. Phys. A 532, 351c (1991)
[$\hat D_{\rm NN}$ given therein is incorrect]

\noindent
6) J.C. Collins, Nucl. Phys. B 396, 161 (1993)

\noindent
7) Xiangdong Ji, CTP\#2219 (MIT, Cambridge, June 1993).

\noindent
8) R.A. Kunne {\it et al}, preprint of Laboratoire National Saturne,
LNS/Th/93-1

\noindent
9) The HELP Collaboration, CERN/LEPC 93-14, LEPC/P7, Sept. 29, 1993.

\noindent
10) Le Yaouanc, LL. Oliver, O. P\`ene and J.-C. Raynal, {\it Hadron Transitions
in the Quark Model}, Gordon and Breach, p. 99.

\noindent
11) B. Andersson, G. Gustafson, G. Ingelman and T. Sj\"ostrand, Phys. Reports
97 (1983).

\noindent
12) X. Artru, J. Czyzewski and H. Yabuki, in preparation.

\noindent
13) D.L. Adams {\it et al}, Phys. Lett. B 264, 462 (1991) ; Z. Phys. C 56, 181
(1992).
\vfill\eject

% Instructions for inclusion of Figures 1, 2, 3 at the bottom of page 4
% [the Postscript file of the figures (to be named " figprot.ps ")
% is appended below, after the \end statement] :

\vglue 3 true cm
\input epsf
\epsfxsize=14 true cm
\centerline{\epsfbox{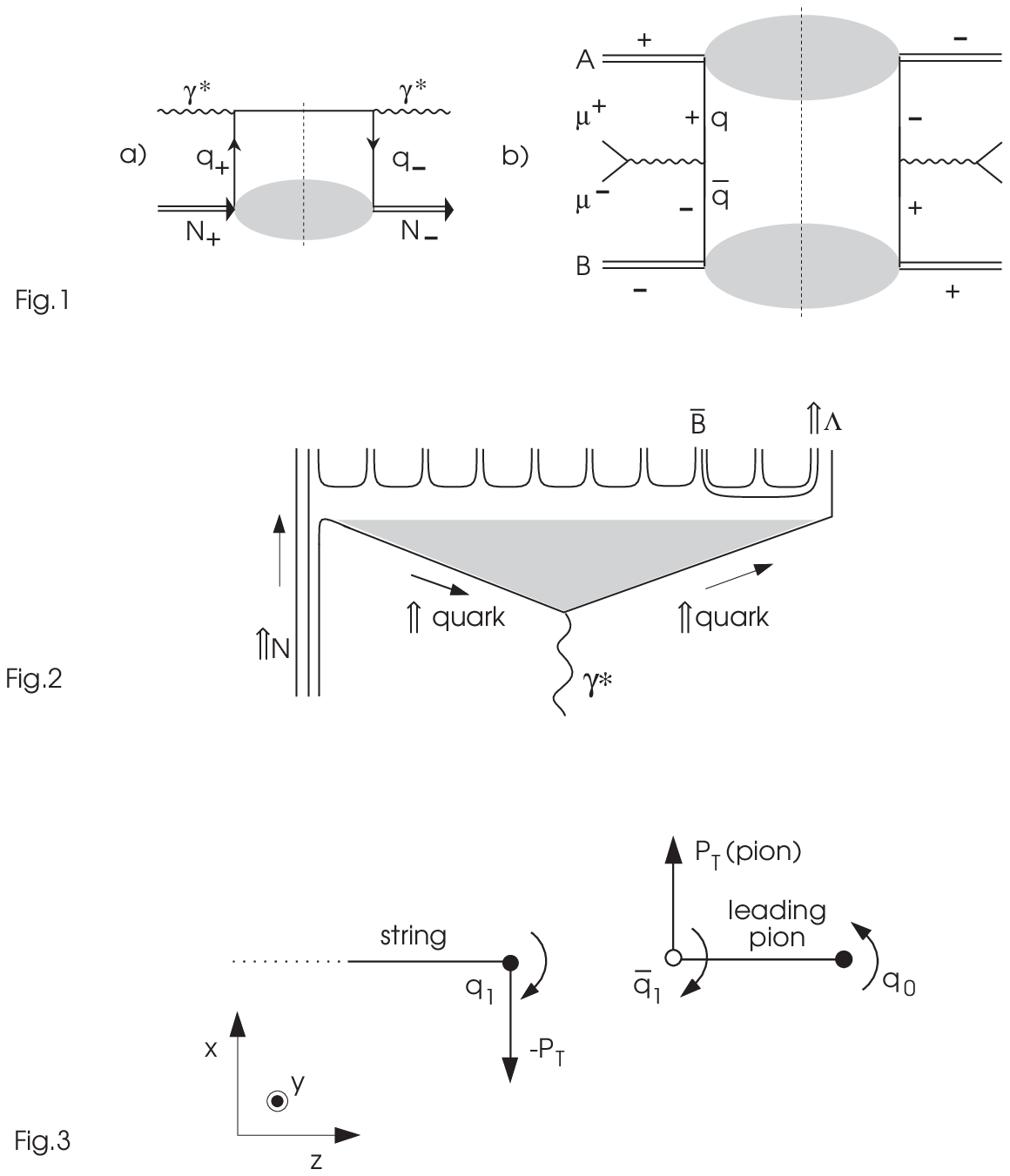}}

\end